\documentclass[prd,twocolumn,amsmath,aps,floats,amssymb, floatfix, superscriptaddress, nofootinbib,10pt]{revtex4-1}
\usepackage{wasysym}
\pdfoutput=1 

\usepackage[T1]{fontenc} 
\usepackage[utf8]{inputenc}
\usepackage{amsthm}
\usepackage{amsmath}
\usepackage{amssymb}
\usepackage{amsfonts}
\usepackage{graphicx}
\usepackage{mathtools}
\usepackage{slashed}
\usepackage{url} 
\usepackage{changepage}
\usepackage{mathrsfs}
\usepackage{float}
\usepackage{bbm}
\usepackage{hyperref}
\usepackage{bm}
\usepackage{rotating}
\usepackage[letterpaper,lmargin=2.5cm,rmargin=2.5cm,tmargin=2.5cm,bmargin=2.5cm]{geometry}
\usepackage{xcolor}
\usepackage{enumitem}
\usepackage[bottom]{footmisc}
\usepackage{svg}
\usepackage{multirow}

\usepackage{tikz}
\usetikzlibrary{math}
\usetikzlibrary{decorations.markings}
\usetikzlibrary{arrows.meta}

\hypersetup{
    colorlinks = true,
    citecolor = {black},
    urlcolor = {blue},
    linkcolor= {darkgray},
}
\DeclareUnicodeCharacter{00A0}{ }

\def\beq{\begin{equation}}
\def\eeq{\end{equation}}
\def\be{\begin{equation}}
\def\ee{\end{equation}}
\def\bea{\begin{eqnarray}}
\def\eea{\end{eqnarray}}

\newcommand{\noun}[1]{\textsc{#1}}

\theoremstyle{plain}
\newtheorem{thm}{\protect\theoremname}
  \theoremstyle{plain}
  
  \theoremstyle{definition}

\usepackage[english]{babel}
  \providecommand{\definitionname}{Definition}
  \providecommand{\propositionname}{Proposition}
\providecommand{\theoremname}{Theorem}

\newtheorem{definition}{Definition}

\begin{document}

\title{On the past-completeness of inflationary spacetimes}
\medskip\
\author{J.~E.~Lesnefsky}%
\email[Email:]{jlesnefs@asu.edu}
\affiliation{Department of Physics,
Arizona State University, Tempe, AZ 85287-1504, USA}

\author{D.~A.~Easson} 
\email[Email:]{easson@asu.edu}
\affiliation{Department of Physics,
Arizona State University, Tempe, AZ 85287-1504, USA}

\author{P.~C.~W.~Davies} 
\email[Email:]{Paul.Davies@asu.edu}
\affiliation{Beyond Center for Fundamental Concepts in Science, Arizona State University, Tempe, Arizona 85287, USA}
\affiliation{Department of Physics,
Arizona State University, Tempe, AZ 85287-1504, USA}

\begin{abstract}
We discuss the question of whether or not inflationary spacetimes can be geodesically complete in the infinite past. 
Geodesic completeness is a necessary condition
for averting an initial singularity during eternal inflation. It is frequently argued that cosmological models which are expanding sufficiently fast (having average Hubble expansion rate $H_{avg}>0$) must be incomplete in null and timelike past directions. 
This well-known conjecture relies on specific bounds on the integral of the Hubble parameter over a past-directed timelike or null geodesic. 
As stated, we show this claim is an open issue. We show that the calculation of $H_{avg}$ yields a continuum of results for a given spacetime predicated upon the underlying topological assumptions. We present an improved definition for $H_{avg}$ and introduce an uncountably infinite cohort of cosmological solutions which are geodesically complete despite having $H_{avg}>0$.  We discuss a standardized definition for inflationary spacetimes as well as quantum (semi-classical) cosmological concerns over physically reasonable scale factors.
\end{abstract}
\maketitle
\noindent

\section{Introduction}
\label{sec:Introduction}
One of the oldest problems of philosophy is whether the universe had an ultimate origin. In the context of relativistic cosmology, that is usually translated into the question of whether there was a past singularity. In standard Friedmann Robertson Walker (FRW) big bang models, causal development of the initial curvature singularity is indeed a past boundary of spacetime. By contrast, the once-popular steady-state cosmological model, involving the continuous creation of particles \cite{HoyleNarlikar1964:1,HoyleNarlikar1964:2,HoyleNarlikar1966}, is eternal. Criticism of the big bang theory by advocates of the steady-state theory were summed up by the aphorism that ‘it merely claims things are as they are because they were as they were.’ In other words, the observed nature of the universe is simply relegated to unexplained initial conditions, and specifically to its ordered, low-entropy, initial state. However, in the steady state model the theory is also incomplete, but for a different reason. The metric for a FRW space is given as:
\begin{equation}
ds^2 = - dt^2 + f^2(t) \left( \frac{dr^2}{1-kr^2}+ r^2 d\Omega^2 \right) \label{FRW metric defn}
\end{equation}
with $f$ being the scale factor and $k$ being the spatial constant sectional curvature, usually normalized to $k \in \{ 0, +1, -1 \}$, and $d \Omega^2$ being the usual spherical metric.  The steady state universe is described as a patch of de Sitter space with $f\left( t \right) = e^{H t}$ and $k=0$.  The steady state universe, which expands at all times, occupies one-half of the complete de Sitter manifold (topologically $\mathbb{R}^1_1 \times S^{n-1}$).  A coordinate system that covers the entire de Sitter spacetime is given by the metric of Eq.~\ref{FRW metric defn} with $f = \mathrm{cosh} \left( H t \right)$ and $k=+1$.  It describes a universe that contracts to a minimum radius and then expands in a time-symmetric manner. Consider a world line $r = \mathrm{constant}$ in the latter system. Tracing back in time from future infinity, the world line traverses the steady state patch and eventually crosses the edge of the coordinate system, and hence out of the steady state universe. Clearly the steady state universe is geodesically incomplete. Recognizing this shortcoming, Hoyle and Narlikar proposed a modified theory \cite{HoyleNarlikar1964:1,HoyleNarlikar1964:2,HoyleNarlikar1966} using the entire spacetime, which involved a contraction followed by an expansion. This so-called C-field cosmology approaches the steady state in the far future and far past and can be made time symmetric by assuming matter creation to be oppositely directed in the regions $t > 0$ and $t < 0$. A closely analogous issue arises in the case of eternal inflation, in which the ‘units of creation’ are not subatomic particles but ‘bubble universes.’ A simple model of eternal inflation uses the steady state metric outside of the bubbles, and is therefore geodesically incomplete.  However, this model can be amended with a minimum spacelike diameter, just like the full de Sitter case, which is complete.

Some confusion arises from the ambiguous use of the term `universe.' One definition is the entire spacetime, including all extensions, e.g. metric $f = \mathrm{cosh} \left( H t \right)$ and $k = +1$ in the foregoing example. Another definition is a congruence of worldlines orthogonal to a spacelike foliation bounded by horizons, e.g. steady state metric in the foregoing. A steady state observer on a worldline $r = \mathrm{constant}$ in this metric perceives an eternally expanding ‘universe’ with an event horizon, but the `universe' (i.e. the timelike foliated coordinate patch shown in Fig.~\ref{fig:light cone}) is past incomplete.

Here we discuss how to characterize the geodesic completeness of a generalization of FRW spacetimes, and show that the aforementioned heuristic example holds in a more general sense.  Our result is purely geometric, holding for any metrical theory of gravity.  

\section{Inflation and Geodesic Completeness}
\label{sec:Inflation and Geodesic C}
Among the most compelling models of the early universe is the inflationary universe paradigm~\cite{Guth:1980zm}.
Shortly after its conception the question naturally emerged: could inflation be eternal, in particular, eternal into the past?  If so, the universe itself could be eternal without the need for a definite beginning \cite{Linde:1986fc,Linde:1986fd,Goncharov:1987ir,Borde:1993xh,Borde2003,Mithani:2012ii}.
A critical component needed to answer
this question is a thorough understanding of singularities in general
spacetimes, see \emph{e.\!\!~g.}~\cite{Hawking1975,Beem1996,ONeill1983}, namely, spacetimes without any further assumptions on structure.

If a spacetime contains a singularity (curvature singularity or otherwise) it cannot inflate eternally. 
One should ascertain whether a spacetime admits even a single \emph{incomplete} geodesic, namely a geodesic which does not exist for all values of affine parameter.  If such an incomplete geodesic exists, the spacetime is singular; an observer on such a timelike geodesic trajectory experiences a catastrophic halt to their proper time, as their worldline cannot be extended past a particular spacetime event. 
In the case of a curvature singularity an observer's clock stops because a spatial defect is encountered.

Venturing forward with pedantic mathematical concern, one must restrict to the class of \emph{maximal} geodesics--curves which cannot be further extended. A maximal geodesic line $\gamma$ is defined as 
\be
\gamma : \left( a , b \right) \rightarrow \mathcal{M} \label{geodesic eqn}
\ee
where $a,b$ is are real numbers in \emph{open} interval $\left( a , b \right)$ and $\mathcal{M}$ is the spacetime in question.  Additionally, one can define a maximal geodesic ray emanating from a point $p = \gamma \left( a \right)$ by
\begin{equation}
    \gamma : \left[ a , b \right) \rightarrow \mathcal{M} \label{geodesic ray eqn}
\end{equation}
The singularity structure of a spacetime may be elucidated by studying the world lines of observers on maximal geodesic trajectories.  If $\gamma$ is complete the domain is defined on all $\mathbb{R}$.  

Now consider a smooth past directed causal geodesic segment $\gamma \left( \left[ a , b \right] \right)$ as the image of a geodesic $\gamma$ defined over the compact set $\left[ a , b \right]$.  Smooth (hence continuous) maps preserve compact sets and therefore $\gamma \left( \left[ a , b \right] \right)$ is compact.  Could $\gamma \left( \left[ a , b \right] \right)$ be complete?  The answer is no.  Either: 
\begin{enumerate}
    \item A singularity is encountered in the image of $\gamma$ and the preimage of such must be excluded from the domain (hence $\left[ a , b \right]$ cannot be all $\mathbb{R}$)
    \item $\gamma \left( a \right)$ and / or $\gamma \left( b \right) \in \partial \mathcal{M}$, the boundary of the spacetime, and is trivially incomplete
    \item $\gamma \left( a \right)$ and / or $\gamma \left( b \right) \in \mathrm{int} \, \mathcal{M}$, the interior of the spacetime, and is extendable
\end{enumerate}
In the first two cases the spacetime is incomplete because any past directed causal geodesic arising from the interior which intersects the spacelike boundary abruptly stops, else it would have to change causal character to propagate along the spacelike boundary which cannot occur for a geodesic.  In the third case, any geodesic segment which realizes an endpoint on the interior can be extended by considering a semi-Riemannian normal neighborhood centered at $\gamma \left( b \right)$ and calculate $\exp_{\gamma \left( b \right)} \dot{\gamma \left( b \right)}$, where $\exp_p X$ is the exponential map centered at $p \in \mathcal{M}$ -- calculated by finding the geodesic at $p$ with initial velocity $X$ and evaluating the point at unit time.  Such an extension must always occur, even if the diameter of such a semi-Riemannian normal neighborhood is very small. Any extendable geodesic segment cannot be complete because it does not contain all the possible information contained in the maximal curve, namely the information contained in the extensions.

In a general model with very
little structure, it can be difficult to classify all geodesics in
a spacetime.~~Even if structure is introduced allowing
such a classification of geodesics, the geodesic equation may not have an analytic solution: thus advanced methods
vis \`{a} vis hyperbolic differential equations and functional analysis
are required.~~However, a sufficient condition for an incomplete
geodesic can be found utilizing Jacobi fields to find conjugate
or focal points, which is often easier to calculate than solving
all possible geodesic equations: this is the basis for the singularity theorems of \cite{Hawking1975}.

Demonstrating that a spacetime is free of incomplete geodesics
does not mean that the spacetime is physically reasonable. While free of singularities, such models may violate reasonable energy conditions with  `exotic' sources required to ensure that all
geodesics are complete.~~Nonetheless, spacetimes
which are geodesically complete are void of singularities at
the geometric level and one may hope that a reasonable metric theory of
gravity exists, capable of supporting these spacetimes.

Returning to the discussion of eternal inflation, the title of \cite{Borde2003} claims that any inflationary spacetime must be geodesically incomplete.~~Below, we provide an uncountably infinite cardnality of \emph{monsters} -- see \cite{Lakatos1976} --
applications of a theory to prickly examples at the boundary of the domain of discourse which helps to clarify tenets of the theory, showing that this cannot be the case.
These simple monsters are well defined within the realm of GR, although in that context,
they violate energy conditions for an arbitrarily short time.~~We begin
with a brief discussion of \cite{Borde2003} elucidating some of the contentious points therein.
We then introduce the class of Generalized Friedmann Robertson
Walker spacetimes (GFRW), and discuss their geodesic completeness.  Next
we discuss properties of scale functions yielding inflationary monsters which satisfy the $H_{avg}>0$
condition of \cite{Borde2003} yet are geodesically
complete.  Finally, we speculate on how quantum gravity may ultimately constrain the domain of discourse with respect to physically reasonable scale functions.
\section{What is Inflation?}
\label{sec:inflation}
Much of the contention surrounding $H_{avg}$ in \cite{Borde2003} stems from ambiguous underlying assumptions.  We remedy this by attempting to explicitly state the underlying assumptions.  Fittingly, we begin by discussing what is meant by the term \emph{inflationary}.

In the introduction of \cite{Guth:2007ng}, inflationary theory pioneer Alan Guth aptly states, ``the term \emph{inflation} encompasses a wide range of detailed theories'', and refrains from defining inflation in a mathematically precise manner. Popular definitions of inflation include ``a quasi-de Sitter period'',  ``a period when gravity acted as a repulsive force'' or ``a period during which the Hubble sphere is shrinking''. Several authors prefer to discuss inflation in terms of \it what it does\rm, avoiding the need for a precise or rigorous definition.  In \cite{Liddle:2000cg}, the authors define inflation as ``any epoch during which the scale factor of the Universe is accelerating'', and provide the defining equation 

\begin{definition}
$
 \mathrm{INFLATION} \Longleftrightarrow \ddot{f} > 0 \,, \label{inflation defn} 
$
\end{definition}
\noindent
where we take $f$ to be the scale factor and ``$\cdot$'' denotes differentiation with respect to cosmic time $t$.
This definition assumes that the metric can be expressed as a warped product from \cite{Bishop1969} as 
\be
g = d \pi_1^* g_1 + f^2 \, d \pi_2^* g_2 \label{warped product defn} 
\ee 
where topologically the manifold in question is a Cartesian product  $\mathcal{M}_1 \times \mathcal{M}_2$ of two semi-Riemannian manifolds $\left( \mathcal{M}_1 , g_1 \right)$ and $\left( \mathcal{M}_2 , g_2 \right)$ and $d \pi_1^*, d \pi_2^* $ are the metric pullbacks of the canonical projections $\pi_1  : \mathcal{M} \rightarrow \mathcal{M}_1$,  $\pi_2 : \mathcal{M} \rightarrow \mathcal{M}_2$ in the Cartesian product construction.  Additionally, $f$ is a strictly positive smooth function ``warping'' the contribution of the second metric $g_2$ to the overall metric $g$.  

Relating this to the standard Friedmann Robertson Walker (FRW) model of
cosmology and cosmogony, one has already foliated out a privileged
time dimension by assumption and by defining a scale function $f$ with respect to this
time dimension. The spacetime metric is that of Eq. \ref{FRW metric defn}.

In fact, the results apropos scale functions discussed in this paper are applicable to the broader class of Generalized Friedmann Robertson Walker (GFRW) spacetimes.  These GFRW spacetimes are defined as the warped product 
\be
\mathbb{R}^1_1 \times_f \Sigma \label{GFRW defn}
\ee
where $\left( \Sigma , g_\Sigma \right)$ is a complete\footnote{Any differentiable manifold admits a Riemannian metric - which can be pulled up from the underlying Euclidean space via the atlas, and any Riemannian metric is conformal to a complete metric~\cite{Nomizu1961}.  Thus, given any (Hausdorff, second countable) reasonable space $\Sigma$, it can be endowed with a complete Riemannian metric.} Riemannian manifold -- taken to be the spacelike leaf -- and $\mathbb{R}^1_1$ is the timelike foliation.  Note that the subscript is used to denote a negative definite index of the metric in $\mathbb{R}^1_1$ as in \cite{Beem1996,ONeill1983}.  Finally a positive definite smooth warping function $f$ is used in the $\times_f$ notation for specification of the warped product metric $g = -dt^2 + f^2 \, g_\Sigma$.  Additionally, the $d\pi_t^*, \, d\pi_\Sigma^*$ are unambiguous and have been suppressed.

One of the goals of \cite{Borde2003} is to generalize concepts from a highly structured FRW spacetime, so one may ask if a spacetime which cannot admit a global foliation of a time direction should be considered inflationary or not.  In fact, if a dimension 3 or higher open neighborhood $U \subset \mathcal{M}$ admits a curl-free, local Killing vector field it is isometric to a warped product spacetime -- see \cite{Beem1996} Lemma 3.78.  Are such spacetimes to be considered inflationary?  In particular, it is much easier for local Killing vector fields to exist than global ones, so many spacetimes which cannot globally foliate out a time direction fall into this category.

An additional concern with respect to Defn.~\ref{inflation defn} are bouncing cosmologies.  Clearly, we must have $\ddot f>0$ for a time during the transition from contraction to expansion; however, the contracting phase strictly violates the definition of \cite{Liddle:2000cg}.  Although strict definitions of a \emph{spacetime} varies subtly between authors, the canonical literature typically demands that the Lorentzian manifold be time oriented, thus it makes sense to talk about $\ddot{f}$ being explicitly positive or negative.  Because there can only be a single choice of time orientation, once this time orientation is fixed what constitutes inflation or deflation is fixed.  Given the above, we feel that it is myopic to exclude the possible bouncing phenomenology from inflationary models and a rigorous definition for inflation should encompass bouncing cosmologies (in the cosmological bounce the co-moving hubble radius decreases, as it does during inflation and there is at least a short period during which $\ddot f >0$ during the bouncing phase.).

Given these considerations, we feel that the inflation definition of Defn.~\ref{inflation defn} is overly restrictive.  We propose a new definition for an inflationary spacetime:
\begin{definition}
Let $\left( \mathcal{M}, g \right)$ be an $n$ dimensional spacetime which admits a connected open neighborhood $U \subset \mathcal{M}$  which is isometric to $ \left( a,b \right) \times_f V$ as a warped product open submanifold, where $\left( a,b \right)$ is a timelike codimension $n-1$ embedded submanifold, $V$ is a spacelike codimension 1 embedded submanifold and no assumptions are made concerning $f$ except that it is a well defined function between sets.  The spacetime $\left( \mathcal{M}, g \right)$ is an \emph{inflationary spacetime} if there exists some $t_0 \in \left( a,b \right)$ such that (assuming that it exists and is well defined), $\ddot{f} \left( t_0 \right) > 0$ where derivatives of $f$ are taken with respect to the timelike $\left( a,b \right)$ coordinate and the sign of $\ddot{f}$ is given with respect to the time orientation of $\mathcal{M}$.  \label{my inflation defn}
\end{definition}
In this definition, no assumptions about the continuity of $f$ or its derivatives are made.  There is debate in the literature about what precise assumptions functions in metrical theories of gravity must have for well defined curvatures to exist -- see \cite{Hawking1975} Chapter 3.1 for example -- we do not comment on this issue here.  Additionally, we do not assume that $f>0$, as in \cite{Bishop1969,Beem1996,ONeill1983}, so as not to exclude the possibility of an initial singularity caused by $f\rightarrow 0$.

There are several technicalities apropos Defn.~\ref{my inflation defn}.  The first is a local global relationship of the local inflationary neighborhood  $ \left( a,b \right) \times_f V$ to the global spacetime $\left( \mathcal{M}, g \right)$.  As written in Defn. \ref{my inflation defn}, the inflationary canonical time axis must be timelike with respect to the global metric and $V$ must be spacelike with respect to this metric as well.  This assumes the existence and causal knowledge of $\left( \mathcal{M}, g \right)$ non-locally.  If one holds the view that inflation is purely a local phenomenon, Defn. \ref{my inflation defn} is overly restrictive, excluding the possibility of two distant observers undergoing inflation orthogonal to each other, each believing the other's time axes to be spacelike - see Fig.~\ref{fig:light cone}.  In fact, upon accepting these arguments one could even theorize that having a well defined metric is a local property!  If one demands causality -- even exotic causality such as in closed timelike curves or vicious spacetimes\footnote{A spacetime is vicious at a point if the causal future / past encompasses the entire spacetime, namely $\mathcal{J}^\pm \left( p \right)= \mathcal{M}$.  A spacetime is totally vicious if it is vicious everywhere -- see \cite{Beem1996,Minguzzi:2006sa}} -- this possibility must be excluded as it is in the definition.  However, by examining various inflationary neighborhoods within a single spacetime under Defn.~2, it gives valuable information as to the causal structure, and subsequently the overall metric of the global spacetime, especially as one extrapolates upon a distant inflationary neighborhood which may exceed one's observable universe.

\newcommand{\drawLightCone}[4]{
    \begin{scope}[shift={(#1,#2)},scale=#3,rotate=#4]
        \draw[thick] (0,0.5) ellipse (0.5 and 0.15);
        \draw[thick] (0,-0.5) ellipse (0.5 and 0.15);
        \draw[thick] (-0.5,0.45) -- (0.5,-0.45);
        \draw[thick] (0.5,0.45) -- (-0.5,-0.45);
    \end{scope}
}

\tikzmath{\nbhdW = 2; \nbhdH=4;}

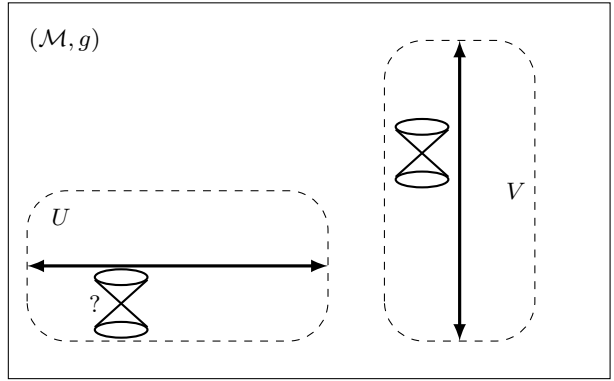
\begin{figure}[ht]
    \centering
    \begin{tikzpicture}
    \coordinate (timelikeNbhd) at (5,0.5);
    \coordinate (spacelikeNbhd) at (0.25,0.5);
    
    \draw (0,0) rectangle (8,5);
    \node at (0.75,4.5) {$\left( \mathcal{M},g \right)$};
    
    \draw[dashed, rounded corners=15] (timelikeNbhd) rectangle +(\nbhdW,\nbhdH) node at +(\nbhdW-.25,\nbhdH/2) {$V$};
    \draw[very thick,latex-latex] (timelikeNbhd) +(\nbhdW/2,0) -- +(\nbhdW/2,\nbhdH);
    \drawLightCone{5.5}{3}{0.7}{0}

    \draw[dashed, rounded corners=15] (spacelikeNbhd) rectangle +(\nbhdH,\nbhdW) node at +(.45,\nbhdW-.35) {$U$};
    \draw[very thick,latex-latex] (spacelikeNbhd) +(0,\nbhdW/2) -- +(\nbhdH,\nbhdW/2);
    \drawLightCone{1.5}{1}{0.7}{0}
    \node at (1.15,1) {?};
    
    \end{tikzpicture}
    \caption{An example spacetime which does not conform to Defn. \ref{my inflation defn}.  The external rectangle is schematic for overall spacetime $\left( \mathcal{M}, g \right)$ which contains two potential inflationary neighborhoods $U,V$.  Neighborhood $V$ has a canonical time axis which is timelike with respect to overall metric $g$ and is indeed inflationary with respect to the above definition.  However, neighborhood $U$ has a canonical ``time'' axis which, in its native immersed Lorentzian (causal) structure is taken to be timelike; however, it is spacelike with respect to overall metric $g$.  Such a situation is excluded by Defn. \ref{my inflation defn}.  Light cones are shown with respect to overall metric $g$.}
    \label{fig:light cone}
\end{figure}

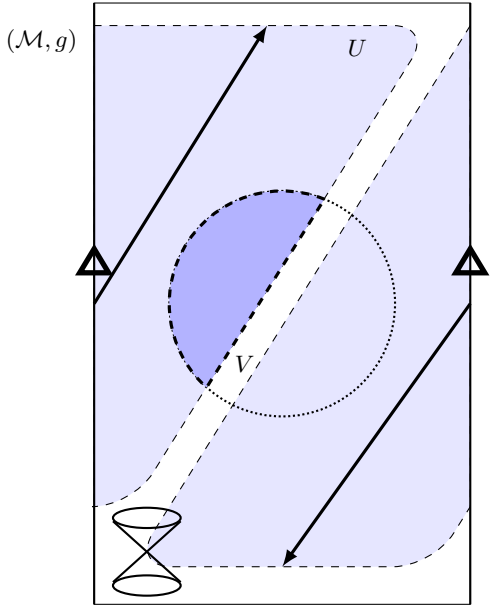
\begin{figure}[ht]
    \centering
    \begin{tikzpicture}
        [decoration={markings, mark= at position 0.6 with {\arrow[scale=2.5]{Triangle[open]}}}] 
    
        \draw (0,0) rectangle (5,8);
        \node at (-0.7,7.5) {$\left( \mathcal{M},g \right)$};

        \draw[fill=blue!10,dashed, rounded corners=15] (5,7.7) -- (0.5,0.5) -- (4.5,0.5) [sharp corners] -- (5,1.3) -- cycle;
        \draw[very thick,-latex] (5,4) -- (20/8,0.5);
        
        \draw[fill=blue!10,dashed, rounded corners=15] (0,1.3) -- (0.5,1.3) -- (4.5,7.7) [sharp corners] -- (0,7.7) -- cycle;
        \draw[very thick,-latex] (0,4) -- (2.3,7.7);
        
        \draw[thick,densely dotted] (2.5,4) circle (1.5);
        \draw[very thick,dashed,fill=blue!30] (3.06,5.39) arc (68:228.0:1.5) -- cycle;
        
        \drawLightCone{0.7}{0.7}{0.9}{0}
        
        \draw[postaction={decorate},thick] (0,0) -- (0,8);
        \draw[postaction={decorate},thick] (5,0) -- (5,8);
        
        \node at (3.5,7.4) {$U$};
        \node at (2.0,3.2) {$V$};
        
    \end{tikzpicture}
    \caption{An example spacetime $\left( \mathcal{M}, g \right)$ conformal to $\mathbb{R}^1_1 \times S^1 \times S^2$.  The vertical perimeter is glued aligning the arrows as is done in a standard quotient topology diagram.  The cylinder admits a single inflationary neighborhood $U$ with canonical time axis wrapped around the cylinder.  In the topology of the overall space one considers a neighborhood $V$ - circle neighborhood in the center, but the intersection with the inflationary subspace topology yields a disconnected subset - one connected component shaded dark.  In the (embedded) overall topology all points in $V$ are ``close'', however in the (immersed) subspace topology the connected components are ``far away''.  Defn. \ref{my inflation defn} excludes this possibility.}
    \label{fig:embedded cyl example}
\end{figure}

We now discuss various subtleties concerning the immersion versus the embedding of the inflationary neighborhood.  Should a spacetime like that of Fig.~\ref{fig:embedded cyl example} be licit under the definition of inflation?  As was mentioned in the previous paragraph, the reader's opinion of local versus global concerns drives possible definitions of inflation.  If one believes inflation to be observer dependent, an immersed submanifold definition is superior and the spacetime of Fig.~\ref{fig:embedded cyl example} is allowed because what occurs outside of the observer's observable universe is absurd.  This has the drawback that every observer could potentially have a different immersed topology.  However, if one believes that the ``top down'' structure of the entire spacetime is paramount, the embedded submanifold definition -- this is what is assumed in Defn.~\ref{my inflation defn} -- and all possible observers must be compatible with the entire spacetime's topology, and Fig.~\ref{fig:embedded cyl example} is excluded.

What about the maximality of inflationary neighborhoods?  As discussed in the introduction, it might be possible to find an extension of an inflationary neighborhood enlarging it.  Thus, the length of a geodesic to the boundary of an inflationary neighborhood can vary depending upon the arbitrary selection of neighborhood boundary.  However, if one chooses to use the \emph{maximal inflationary neighborhood} and a geodesic intersects the boundary in finite length then the space is incomplete.  Defn. \ref{my inflation defn} does not comment on this phenomena, and it is possible one could be interested in inflationary phenomena in a particular coordinate system without wanting maximality.  However, if one desires to compute geodesic completeness in an inflationary neighborhood, it must be maximal.

Does an inflationary neighborhood need to be connected?  The explicit use of the term ``connected'' in Defn. \ref{my inflation defn} is equivalent to the connectedness of the spatial section $V$ in the Cartesian product construction of $\left( a , b \right) \times_f V$.  In order for the definition to correlate with observation it is necessary that the inflationary neighborhood be connected: how would one observe a distinct\footnote{The use of the term ``distinct'' implies each spatial foliation is disjoint from every other one.} distant connected component?  However, in some theories - in particular those which admit pocket universes - it is possible that inflation is occurring in multiple distinct distant places at once, and as long as the geometry of the inflating regions is uniformly scaled by the same $f$ the inflationary patch could be disconnected (topologically $\left( a , b \right) \times_f \coprod_{i \in \mathcal{I}} V_i$ for some index set $\mathcal{I}$).  One could consider the union of all inflating regions - even those with differing scale functions - as a union: this could be useful in calculations of inflating region volumes \cite{Borde2003,Borde:1993xh,Mithani:2012ii}.  All of the aforementioned concerns are built on the edifice of Defn. \ref{my inflation defn}; the definition itself, however, only defines the geometry of a single connected inflating region akin to our observable universe.

Finally one can consider the dimension of the spacelike foliation $V$.  In standard metric theories of gravity, this must be a codimension 1 submanifold, akin to a hypersurface.  If, however, one would like to accommodate a theory with extra dimensions, it is conceivable that other codimensions might be considered, perhaps for various D-branes populating the bulk.  This is not accommodated by Defn. \ref{my inflation defn} in order to agree with FRW historical convention. It would not be difficult to alter the definition to include this phenomenology.

Defn.~\ref{my inflation defn} broadly expands the phenomenology which is considered inflationary.  In the absence of a unique model of the nascent universe supported by experimental evidence, we choose to be overly accommodating with respect to the inflationary paradigm, which is consistent with known observations.  Additionally, there is precedent with defining a categorization of a spacetime with respect to a quantity at a single point -- consider the general condition of \cite{Hawking1975}.  One my also discuss how much inflation a spacetime is experiencing by using statements like it is ``completely inflationary'' where it inflates for all time, or only inflates on $\left( a,b \right)$, to accommodate realistic models and also bouncing models.

We adopt Defn.~\ref{my inflation defn} as a working definition for inflation, attempting to address some of the topological machinery enthymeme in Defn.~\ref{inflation defn}.  Naturally, depending on one's bias apropos phenomenology, there are other potential definitions for inflation which might be superior.  For example, the topological criteria for an inflationary neighborhood to exist -- namely the existence of a curl free local Killing vector field -- might uniquely specify a definition upon further investigation and proof. We leave such exploration for future work.

\section{Discussion of $H_{avg}$}
\label{sec:h_avg disc}
We now compare our proposed definition for an inflationary spacetime,  Defn. \ref{my inflation defn} with a discussion of $H_{avg}$ in \cite{Borde2003}.  There, the authors propose a gedanken experiment of measuring the tidal forces in a
general spacetime $\left(\mathcal{M},g\right)$ along a geodesic $\gamma$ with velocity
$V^{\mu}=\frac{\mathrm{d}\gamma}{\mathrm{d}\lambda}$ by subsequent
measurements of a massive test particle moving along some timelike
geodesic $\alpha\,:\,J\rightarrow\mathcal{M}$ with velocity
$U^{\mu}=\frac{\mathrm{d}\alpha}{\mathrm{d}\lambda}$ in a locally Minkowski neighborhood.~~The quantity
\begin{equation}
H^{\gamma} = -\frac{V_{\mu}\left(\lambda\right)}{\beta^{2}\left(\lambda\right)+\kappa}\frac{\mathrm{D}U^{\mu}\left(\lambda\right)}{\mathrm{D}\lambda}\label{eq:H BGV}
\end{equation}
 along geodesic
$\gamma$, where a spatially dependent generalized Lorentz factor
is given by $\beta^{2}\left(\lambda\right)=V_{\mu}U^{\mu}$, and $\kappa=V_{\mu}V^{\mu}$
is taken to be $\kappa=-1$ for $\gamma$ being a timelike geodesic, and $\kappa=0$
for $\gamma$ being a null geodesic.\footnote{We use metric signature convention $\left\{ -1,+1,\cdots,+1\right\} $ opposed to
$\left\{ +1,-1,\cdots,-1\right\} $ used in \cite{Borde2003}.}~~Additionally,
we denote the parallel transport of $U^{\mu}$ along $\gamma$ as
$\frac{\mathrm{D}U^{\mu}}{\mathrm{D}\lambda}$.  The above may be written succinctly in geometers notation as
\begin{equation}
    H^{\gamma} = - \frac{g \left( V , \nabla_{V} U \right)}{g \left( U , V \right) + \kappa} \label{eq:H BGV geometer}
\end{equation}
where $g$ is the metric and $\nabla$ is the Levi-Civita connection.

Note that Eq.~\ref{eq:H BGV} 
reduces to the usual definition of the Hubble parameter
\begin{equation}
H\left(t\right)=\frac{\dot{f}\left(t\right)}{f\left(t\right)}\label{eq:H FRW}
\end{equation}
in the special case that the spacetime is inflationary, and in the inflationary neighborhood, $V^\mu$ is purely spacelike and $U^\mu$ is purely timelike\footnote{This calculation makes use of Corollary 12.8 in \cite{ONeill1983}.}.  In fact, in a gedanken experiment where timelike observer $V = a \frac{\partial}{\partial t} + A$ measures a timelike test particle $U = b \frac{\partial}{\partial t} + B$ with both $A , B$ purely spacelike, we find
\be
    H^\gamma = - \frac{b \frac{\dot{f}}{f} \lVert A \rVert ^2_g - a^2 \frac{\partial b}{\partial t} + g \left( A , d \pi_\Sigma \nabla_A B \right) }{g \left( A , B \right) - ab - 1} \label{havg calc full}
\ee
where $d \pi_\Sigma \nabla_A B $ is the projection of the $\Sigma$-tangential components of $\nabla_A B$ onto spacelike slice $\Sigma$ and $\lVert A \rVert^2_g = g \left( A , A \right)$.  By utilizing the fact that in warped product constructions $\xi^2_A \equiv f^2 \lVert A \rVert^2_g$ is a constant -- which corresponds to the initial speed of a test particle with respect to cosmic time, not a N{\"o}ether conserved quantity -- Eq. \ref{havg calc full} can be expressed as
\begin{widetext}
\begin{equation}
    H^\gamma = \frac{\dot{f}}{f} \left( \frac{\sqrt{1+\frac{\xi^2_B}{f^2}}\left( \frac{\xi^2_A}{f^2} \right) + \left( 1 + \frac{\xi^2_A}{f^2} \right) \frac{\xi^2_B}{f^2} \frac{1}{\sqrt{1+\frac{\xi^2_B}{f^2}}} }{1 + \sqrt{1+\frac{\xi^2_A}{f^2}}\sqrt{1+\frac{\xi^2_B}{f^2}} - g \left( A , B \right)} \right) + \frac{g \left( A , d \pi_\Sigma \nabla_A B \right)}{1 + \sqrt{1+\frac{\xi^2_A}{f^2}}\sqrt{1+\frac{\xi^2_B}{f^2}} - g \left( A , B \right)} \label{havg calc final}
\end{equation}
\end{widetext}
which differs from the expected Hubble parameter $\frac{\dot{f}}{f}$.  If we assume spatial curvature is negligible - namely $d \pi_\Sigma \nabla_A B \simeq 0$ and the directions of spatial propagation between observer and test particle are orthogonal, the above equation simplifies to
\be
    H^\gamma \simeq \frac{\dot{f}}{f} \left( \frac{\sqrt{1+\frac{\xi^2_B}{f^2}}\left( \frac{\xi^2_A}{f^2} \right) + \left( 1 + \frac{\xi^2_A}{f^2} \right) \frac{\xi^2_B}{f^2} \frac{1}{\sqrt{1+\frac{\xi^2_B}{f^2}}} }{1 + \sqrt{1+\frac{\xi^2_A}{f^2}}\sqrt{1+\frac{\xi^2_B}{f^2}}} \right) \label{havg calc final slow}
\ee
which is schematic to
\begin{equation}
    H^\gamma \simeq \frac{\dot{f}}{f} \times \mathcal{O} \left( \xi^2_A , \xi^2_B \right) \label{havg calc final schematic}
\end{equation}
Hence, we arrive at the undesirable result, that $H^\gamma$ is irreducibly convoluted with the particular speeds $\xi^2_A, \xi^2_B$ of the test particles in question.  It would be ideal to find an expression which yields $\frac{\dot{f}}{f}$ independent of the particular speeds of the particle, which Eq. \ref{eq:H BGV} is not.  However, in the case that both particles are relativistic with respect to cosmic time one recovers $H^\gamma \simeq \frac{2 \sqrt{2}}{3} \frac{\dot{f}}{f}$, but for non-relativistic particles this pre-factor varies by choice of test particle.

It would be ideal to calculate the Hubble parameter $\frac{\dot{f}}{f}$ in an inflationary spacetime independent of frame and particular velocity.  One proposal is the quantity
\begin{equation}
    H^\gamma = \frac{g \left( \frac{\partial}{\partial t} , \mathrm{II} \left( X , Y \right) \right)}{g \left( X , Y \right)}
    \label{H II}
\end{equation}
where $X,Y$ are purely spacelike (horizontal) vector fields and $\mathrm{II} \left( X , Y \right) = \mathrm{nor} \nabla_X Y$ is the second fundamental form yielding extrinsic curvature, where $\mathrm{nor} \nabla_X Y$ is the projection of $\nabla_X Y$ onto the linear subspace normal to the hypersurface in question.  In a GFRW
\begin{equation}
\mathrm{II} \left( X , Y \right) = - \frac{\dot{f}}{f} g \left( X , Y \right) \frac{\partial}{\partial t}
\label{II GFRW}
\end{equation}
thus, for any $X,Y$ as stated above, one reaps $H^\gamma = \frac{\dot{f}}{f}$.  There are of course drawbacks to this definition: ambiguity of precisely knowing $\frac{\partial}{\partial t}$ or that $\mathrm{II}$ is an abstract geometric quantity and $X,Y$ are spacelike; neither of which can be directly measured.

We now state what we consider to be the main result of
\cite{Borde2003}:
\begin{thm} \label{thm:bgv} 
\emph{\noun{Inflationary spacetimes are past
incomplete}} - Let $\left(\mathcal{M},g\right)$ be a spacetime, and
$\gamma\,:\,\left[\lambda_{i},\lambda_{f}\right]\rightarrow\mathcal{M}$
be some causal geodesic.~~If one computes the quantity 
\begin{equation}
H_{avg}^{\gamma}=\frac{1}{\lambda_{f}-\lambda_{i}}\int_{\lambda_{i}}^{\lambda_{f}}H^{\gamma}\left(\alpha\right)d\alpha\label{eq:havg bgv}
\end{equation}
to be strictly positive along the image of $\gamma$, the spacetime
is geodesically incomplete.
\end{thm}

We study the implications of this theorem and
its results.  First we will examine topological subtleties calculating\footnote{In Thm \ref{thm:bgv} the definition of $H^\gamma_{avg}$ includes the superscript `$\gamma$' to reinforce the idea that $H_{avg}$ is potentially geodesic dependent.  In the case where choice of geodesic is unambiguous the superscript will be supressed.} $H_{avg}$ and introduce two other methods to calculate this quantity including the asymptotic past and / or futures.  Next we will enumerate an uncountably infinite cardinality of classical monsters with respect to Thm.~\ref{thm:bgv} which are geodesically complete despite having various values of $H_{avg} $ depending on the underlying topological assumptions.  Some of the calculated $H_{avg} > 0$ are strictly positive which violates Thm \ref{thm:bgv}.  

Here we consider only metrical theories of gravity;
and while cursory computations of the Einstein Field Equations (EFE)
will be discussed within the context of energy conditions, all proofs that follow
require only assumptions at the \emph{geometric} level.~~Given this, however, our aforementioned GFRW spacetime monsters do exist
within the scope of general relativity, with only an arbitrarily small
necessity for energy-condition-violating exotic matter.  On the other side, these completeness results hold for any metrical theory of gravity which may ultimately preempt general relativity.~~Any further speculations on relationships
between geometrical quantities and mass-energy is beyond the scope of this paper.

Our first criticism of Eq.~\ref{eq:havg bgv}, involves the implicit assumption that the integral
is computed over a \it compact  \rm interval.  Careful study of \cite{Borde2003}  Eqs. 5, 10, 11 reaps that the computed integrals are evaluated on the boundary per Stokes' Theorem\footnote{Here we invoke the full Stokes Theorem $\int_\mathcal{M} d \omega = \int_{\partial \mathcal{M}} \omega$ not the specific application seen in surface geometry.} and thus the underlying topology of the $H_{avg}$ calculation is a compact set, seeing as no mention is made of any limiting procedure not including the boundary.  In this way,
the integral is predestined to converge, assuming the minimum condition that
$H^{\gamma}\left(\lambda\right)$ is never infinite on a set of non-zero
(Borel) measure.~~Canonically,
properties of geodesic completeness are formulated with respect
to \it maximal  \rm geodesic rays of domain $\left[0,b\right)$.  A strict reading of Thm. \ref{thm:bgv} then begs the question of what underlying interval to select, and the value of $H_{avg}$ varies depending on the interval.  No matter which compact interval is chosen, any compact geodesic segment is necessarily incomplete by the discussion from Sec. \ref{sec:Inflation and Geodesic C}.  By defining geodesics in this manner over compact intervals it is difficult to compare Thm. \ref{thm:bgv} to the standard mathematical literature, and even worse is equivalent to incompleteness.  A strict application of Thm. \ref{thm:bgv}, exactly as written, never allows for the possibility that an inflationary spacetime could be complete.

Addressing these criticisms we propose two amended definitions of $H_{avg}$
\begin{equation}
H^{-,\gamma}_{avg}=\lim_{a \rightarrow -\infty}\frac{1}{b-a}\int_{a}^{ b}H^{\gamma}\left(\alpha\right)d\alpha\label{eq:Havg asym past}
\end{equation}
and
\begin{equation}
H^{\pm,\gamma}_{avg}=\lim_{-a,b\rightarrow\infty}\frac{1}{b-a}\int_{a}^{ b}H^{\gamma}\left(\alpha\right)d\alpha\label{eq:Havg asym full}
\end{equation}
The first, $H_{avg}^-$, includes the asymptotic past and the latter $H^\pm_{avg}$ includes the entire worldline including both the asymptotic past and future.  With respect to the maximality requirements of geodesic completeness, the proposed definitions of Eqs. \ref{eq:Havg asym past}, \ref{eq:Havg asym full} remove the topological ambiguity of interval selection inherent in Thm. \ref{thm:bgv}.  Geodesic ray candidates for past complete geodesics correspond to $H^-_{avg}$ and the maximal curves as candidates for a complete geodesic line correspond to $H^\pm_{avg}$ .

~~In particular,
if one defines $H^{\gamma}$ by Eq.~\ref{eq:havg bgv},
the suppression by a finite $\frac{1}{\lambda_{f}-\lambda_{i}}$ makes it
difficult for $H_{avg}\rightarrow0$ in the limit, yielding
false positives for geodesic incompleteness: this is equivalent to an a priori assumption of incompleteness.~~Considering
this, it is not surprising the aforementioned use of $H_{avg}$ claims to find all inflationary spacetimes to be incomplete.  However, if $H^-_{avg}$ is utilized the finite contribution from the start of the geodesic ray, $\gamma \left( b \right)$, is quenched in the limit $a \rightarrow -\infty$ and if Eq. \ref{eq:Havg asym past} is substituted for Eq. \ref{eq:havg bgv} in Thm. \ref{thm:bgv}, one finds $H^-_{avg} = 0$ in many cases -- with the notable example of a cofinite tail of $f = \exp{H t}$ which is already well-known to be incomplete (see Example 7.41 in \cite{ONeill1983}) -- and Thm. \ref{thm:bgv} almost never applies.  This occurs for examples such as the monster of Figs. \ref{GFRW counter-example}, \ref{fig:eternally inflation}  which are non-decreasing and inflationary, and yet $H^-_{avg} = 0$.  Interestingly, if one includes the asymptotic future and calculates $H^\pm_{avg}$, the cofinite deSitter phase restores the support needed for $H^\pm_{avg} > 0$ and such an inflationary spacetime has a positive averaged Hubble parameter as expected, however this calculation is highly sensitive to the order the limits are taken in.

Another interesting subtext of Thm.~\ref{thm:bgv} is the dependence of $H_{avg}^{\gamma}$ on a
particular choice of geodesic $\gamma$.~~In
particular, the geodesic which one selects might not happen
to yield a $H_{avg}^{\gamma}>0$: in order to properly classify geodesic completeness
of $\mathcal{M}$ one must examine $\sup_{\gamma\in\mathrm{MGeo}\left(\mathcal{M}\right)}H_{avg}^{\gamma}$,
where $\mathrm{MGeo}\left(\mathcal{M}\right)$ is the space of maximal
geodesics over $\mathcal{M}$.  Utilizing methods of analysis and the fact that the space of complete maximal geodesics is closed in the $\mathcal{C}^0$ Whitney topology -- see Theorem 4.7 in \cite{Sanchez1998} -- one should be able to calculate the range of $H^\gamma_{avg}$ when viewed as a functional on the space of geodesics over a (fixed) spacetime, which we leave to future work.  Finally, the fact that both the observer and test particle are assumed to follow geodesics has not been used in the derivation of $H^\gamma$.

It is obvious that the $H_{avg}$ of Eqs. \ref{eq:havg bgv}, \ref{eq:Havg asym past}, \ref{eq:Havg asym full} is not linear.  Having some generalization of linearity for $H_{avg}$ would be ideal.  A geodesic segment $\gamma : \left[ \lambda_0 , \lambda_\ell \right] \rightarrow \mathcal{M}$ could be decomposed into $\left[ \lambda_0 , \lambda_\ell \right] = \bigcup_{k=0}^{\ell - 1} \left[ \lambda_k , \lambda_{k+1} \right]$ with $\gamma_k = \gamma \left( \left[ \lambda_k , \lambda_{k+1} \right] \right)$.  One computes
\begin{equation}
    H^\gamma_{avg} \ne \sum_{k=0}^{\ell - 1} H^{\gamma_k}_{avg} \label{eq:Havg sum}
\end{equation}
Without any assumptions about the partition $\{ \left[ \lambda_k , \lambda_{k+1} \right] \}_{k=0}^{\ell -1}$ one cannot bound $H^\gamma_{avg}$ above or below $\sum_{k=0}^{\ell - 1} H^{\gamma_k}_{avg}$, especially if the interval length is not constant.  We return to this issue later with respect to the monsters discussed in the next section.

\section{Monsters}
\label{sec:counter-examples}
We now present an uncountably infinite class of classical monsters which have $H_{avg} \geq 0$ but are geodesically complete.  We would like to remind the reader that this a purely geometric result: it applies to any metrical theory of gravity.  For any GFRW $\mathbb{R}^1_1 \times_f \Sigma$ with a smooth scale function $f$ -- thus guaranteeing there will not be any curvature singularities -- the geodesic completeness of GFRWs is enumerated by the following \cite{Sanchez1994,Sanchez1998}:

    \begin{thm}{Geodesic Completeness Criterion for GFRW Spacetimes - }
        Let $\mathcal{M} = \mathbb{R}^1_1 \times_f \Sigma$ be a GFRW spacetime
        
        \begin{enumerate}
            \item The spacetime $\mathcal{M}$ is future timelike complete iff $ \int_{t_0}^\infty \frac{f \left( t \right) dt}{\sqrt{\left( f \left( t \right) \right)^2 + 1}}$ diverges for all $t_0 \in \mathbb{R}$. 
            \item The spacetime $\mathcal{M}$ is future null complete iff $ \int_{t_0}^\infty f \left( t \right) dt$ diverges for all $t_0 \in \mathbb{R}$. 
            \item The spacetime $\mathcal{M}$ is future spacelike complete iff $\mathcal{M}$ is future null complete or if $\int_{t_0}^\infty f \left( t \right) dt < \infty$ then $f$ is unbounded; for all $t_0 \in \mathbb{R}$. 
            \item The GFRW is past timelike / null / spacelike complete if, for items 1-3 above, upon reversing the limits of integration from $\int_{t_0}^\infty$ to $\int_{-\infty}^{t_0}$ the word ``future'' is replaced by ``past''.
            \item The spacetime $\mathcal{M}$ is \emph{geodesically complete} iff it is both future and past timelike, null, and spacelike geodesically complete.
        \end{enumerate}

        \label{sanchezthm}        

    \end{thm}
Thus, one can definitively calculate the geodesic completeness of a given GFRW with Thm.~\ref{sanchezthm} and compare the predictions to Thm. \ref{thm:bgv}.  In particular, if the condition that 
\begin{equation}
  \inf_\mathbb{R} f = a > 0
\label{f inf} 
\end{equation}
holds then the GFRW spacetime is geodesically complete.  One can imagine a class of GFRW spacetimes with a smooth scale function obeying $\inf_\mathbb{R} f = a > 0$, $\lim_{t \rightarrow -\infty} f = c$ with $c \geq a > 0$, however for $t>t_0$ with respect to some given $t_0 \in \mathbb{R}$ behaves as de Sitter space with $f = e^{Ht}$.  Such scale functions can be constructed with smooth partitions of unity.  

As discussed in Sec. \ref{sec:h_avg disc}, the generalization of the Hubble parameter to Eqs. \ref{eq:H BGV}, \ref{eq:H BGV geometer} yields Eq. \ref{havg calc final} which differs from the expected $\frac{\dot{f}}{f}$.  However, in a GFRW one can always choose a spacelike geodesic measuring a timelike test particle for which $H^\gamma$ does in fact reduce to the expected value.  For the remainder of this section, unless otherwise noted, we will make such a choice that $H^\gamma = \frac{\dot{f}}{f}$.

Now one must examine the question of $H_{avg}$ with respect to these monsters. If one calculates $H_{avg}$ directly from Thm. \ref{thm:bgv} one must determine an interval to calculate over.  Ignoring the fact that any compact interval cannot parameterize a maximal, hence complete, geodesic one may still calculate $H_{avg}$ over the selected interval; however, the value of $H_{avg}$ varies depending on the interval selected: if an interval that only contains the de Sitter phase is selected one yields $H$, but as $f \rightarrow c$ the value of $H_{avg}$ is quenched by the support of the cofinite past towards but never attaining zero.  Depending on the behavior of $f$ in the finite past as well as the interval selected, $H_{avg}$ may obtain any real value.  However, as more of the cofinite de Sitter phase is included in an interval, any bounces are offset by the increasing exponential function, and Thm. \ref{thm:bgv} applies.

\begin{figure*}[ht]
\centering \includegraphics[width=6in]{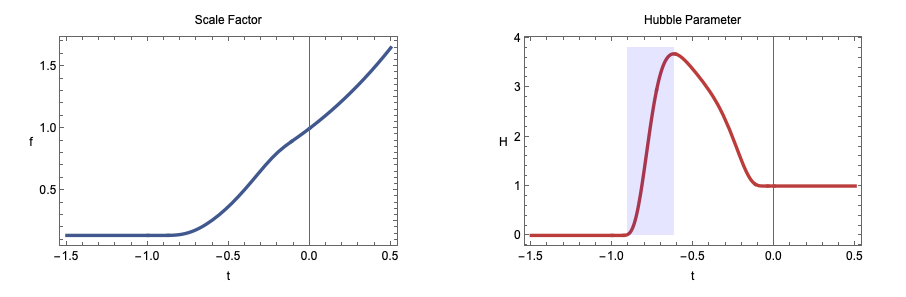}
\caption{Plot of the scale function for a monster GFRW $\mathbb{R}^1_1 \times_f \Sigma$. Here $f$ 
loiters in a Minkowski phase for $t < -1$ and enters a de Sitter phase for $t \geq 0$.  The transition zone, for $t \in \left[ -1 , 0 \right]$, consists of both phases being convoluted with a smooth bump function. By careful construction the spacetime model is
free of curvature singularities commensurate with the smoothness of $f$ and the completeness of $\left( \Sigma , g_\Sigma \right)$. The GFRW
is geodesically complete as per Thm.~\ref{sanchezthm}.  One calculates $H_{avg} \in \left(0, H \right]$, $H^-_{avg} = 0$, and $H^\pm_{avg} \in \left[0 , H \right]$. On the right, the shaded region where $\dot{H} \geq 0$, indicates the NEC (and WEC) is violated.
\label{GFRW counter-example}
}
\end{figure*}

The calculation of $H^-_{avg} = 0$ proceeds as in the previous paragraph, except the asymptote of zero is actually realized.  This is particularly concerning: one such monster in this cohort is that of Fig. \ref{GFRW counter-example}.  In fact, the monster $\exp{H t} + c$ with $c > 0$ (see Fig.~4) inflates eternally but has $H^-_{avg} = 0$.  If $H^-_{avg}$ is calculated in this fashion, Thm. \ref{thm:bgv} does not apply to any monsters in this cohort.

If one selects a past directed geodesic ray with starting point extremely far but still finite in the future -- say $\Sigma \left( 216 \right)$ where $\Sigma \left( n \right)$ is the busy beaver function of \cite{Rado1962} -- the observable universe of such an observer\footnote{Observable universe radius is taken to be the speed of light divided by the Hubble constant.  This also assumes that $H$ of the model is commensurate to the $H$ of the physical universe such that $\Sigma \left( 216 \right) \cdot H \ggg 1$, thus an unfathomable number of $e$-folds have occurred} is only privy to the de Sitter phase and would measure $H_{avg} = H$, however the truth is that there was a cofinite loiter into the asymptotic past and the ``correct'' measurement of $H^-_{avg} = 0$.  Such a universe inflates into the cofinite asymptotic future in a de Sitter phase and yet has $H^-_{avg} = 0$: a paradoxical result.  A criticism of this argument is that a universe in the epoch of $\Sigma \left( 216 \right)$ could be a very strange place, perhaps one that does not support life if an anthropomorphic principle is invoked.  It might be possible that a clever experiment utilizing novel physics might be able to detect the loitering phase at a starting point of only a few Hubble times, or perhaps not.  Pragmatically one must consider empirical and epistemological considerations of various observers piercing various horizons to be able to measure the scale function $f$ and calculate $H_{avg}$.  We have tacitly assumed that global knowledge of $f$ can always be measured and communicated but this may not be true considering the existence of horizons such as the observable universe in a de Sitter phase.  The $\Sigma \left( 216 \right)$ model above features an extreme praxis which calls such an assumption of global knowledge of $f$ -- and thus the applicability of $H_{avg}$ -- into question.  Any further discussion on this matter is beyond the scope of this paper.  

Finally, the calculation of $H^\pm_{avg}$ is even more concerning.  The outcome is predicated upon what order the limits of Eq. \ref{eq:Havg asym full} are taken.  If one executes the $b$ limit first $$\lim_{a\rightarrow -\infty} \lim_{b \rightarrow \infty} \frac{1}{b-a} \int_a^b H^\gamma \left( \alpha \right) d \alpha = H$$ and if the $a$ limit is executed first $$\lim_{b\rightarrow \infty} \lim_{a \rightarrow -\infty} \frac{1}{b-a} \int_a^b H^\gamma \left( \alpha \right) d \alpha = H^-_{avg} = 0$$  Additionally for any $q \in \mathbb{R}^+_*$, the strictly positive real numbers, one can parameterize the $H^\pm_{avg}$ calculation as $$ \lim_{b \rightarrow \infty} \frac{1}{b \cdot \left( q + 1 \right)} \int_{-b}^{qb} H^\gamma \left( \alpha \right) d \alpha\in \left( 0 , H \right) $$ which varies by choice of $q$.  This is particularly concerning, because if $H^\pm_{avg} \in \left[ 0 , H \right]$ Thm. \ref{thm:bgv} both applies and does not simultaneously to the monsters in this cohort.

All that matters to construct this class of counter examples is that the tail of $f$ behaves as de Sitter, and it can be smoothly combined with any other function of positive infimum on the interval $\left( -\infty , t_0 \right)$ with a smooth partition of unity over a compact interval, and approaches a strictly positive constant in the distant past. The size of the function space over this interval enumerates a cohort of classical monsters, see Fig.~\ref{GFRW counter-example}.  Additionally, because the dimensionality of $\mathcal{C}^\infty \left( \left( -\infty , t_0 \right) \right)$, the space of smooth functions on the interval  $\left( -\infty , t_0 \right)$, is uncountably infinite, there are an uncountably infinite number of GFRW spacetimes which have $H_{avg} \geq 0$ but are geodesically complete.

\begin{table}[ht]
    \centering
    \begin{tabular}{|c|c|c|}
        \hline
        \multicolumn{3}{|c|}{Monster Cohort $H_{avg}$ Calculation} \\
        \hline \hline
         $H_{avg}$     & $ \mathbb{R}$ & varies by interval choice and scale factor \\
         \hline
         $H^-_{avg}$   & 0 & $\;$ \\
         \hline
         $H^\pm_{avg}$ & $ \left[ 0, H \right] $ &  varies by limit order and parameterization \\
         \hline
    \end{tabular}
    \caption{$H_{avg}$ calculations for the uncountable cohort of smooth monsters with $\inf_\mathbb{R} f = a > 0, \lim_{t \rightarrow -\infty} f = c \geq a > 0$ and $f = e^{Ht}$ for $t>t_0$.  Depending if $H_{avg}$ is calculated by Eqs. \ref{eq:havg bgv}, \ref{eq:Havg asym past}, \ref{eq:Havg asym full} the answer varies.  The calculation of $H_{avg}$ by Eq. \ref{eq:havg bgv} varies by choice of interval.  The calculation of $H^-_{avg}=0$ with certainty.  The calculation of $H^\pm_{avg}$ varies by parameterization of $\int_a^b H^\gamma \left( \alpha \right) d \alpha$ and the order the limits are taken.  All monsters of this cohort are geodesically complete by Thm. \ref{sanchezthm}.}
    \label{tab:various havgs summary}
\end{table}

\begin{figure*}[ht]
    \centering
    \includegraphics[width=6in]{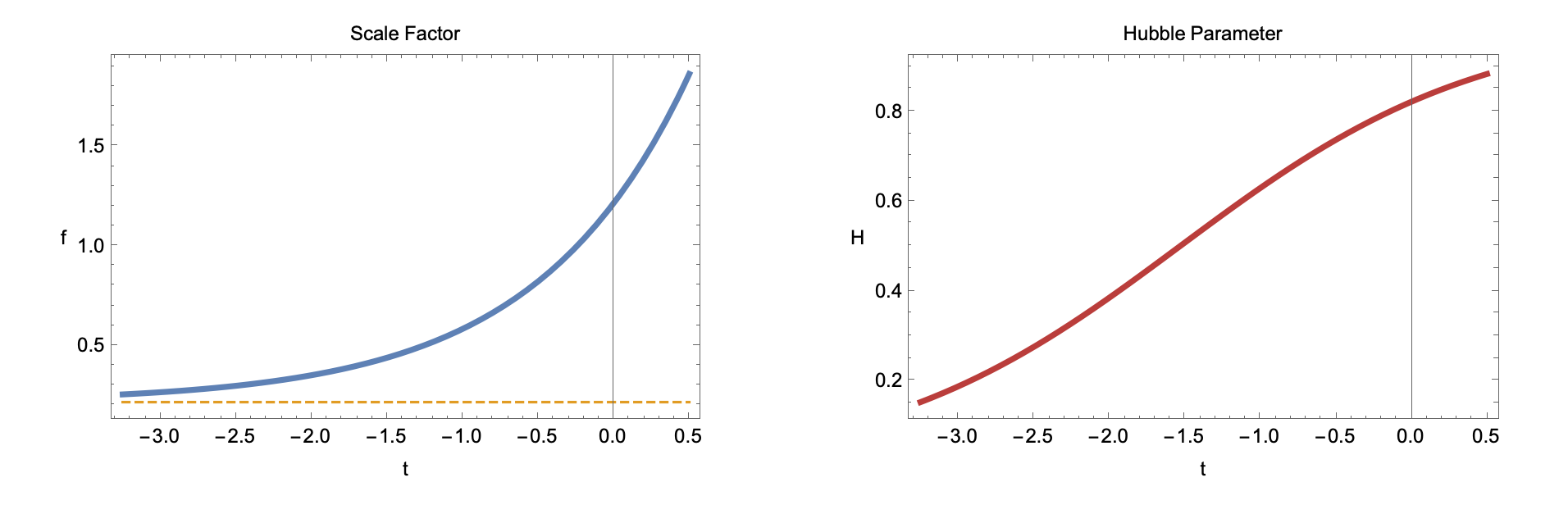}
    \caption{The scale factor for an eternally inflating example GFRW $\mathbb{R}^1_1 \times_f \Sigma$ with $f = \exp{H t} + c$ with $H=1$ and $c=\frac{27}{125} \approx 0.216$.  Strictly speaking, this example is not an element of the uncountable cohort of monsters discussed in this paper due to the additive constant $c$, but Thms. \ref{thm:bgv}, \ref{sanchezthm} still apply.  The scale function is shown in blue with the asymptote shown in dotted orange.  This example is geodesically complete by Thm. \ref{sanchezthm} with $\inf_{\mathbb{R}} f = \frac{27}{125}$.  The various $H_{avg}$ are calculated as $H_{avg} \in \left( 0 , \frac{\exp{H t_f}}{\exp{H t_f} + c} \right)$ as a function of interval, $H^-_{avg} = 0$, and $H^\pm_{avg} \in \left[ 0 , H \right]$.  Thm. \ref{thm:bgv} applies because $H_{avg} > 0$.  Despite being eternally inflating $H^-_{avg} = 0$ and one potential calculation of $H^\pm_{avg} = 0$ as well, a paradoxical result.}
    \label{fig:eternally inflation}
\end{figure*}

One can apply the theory developed here to the loitering monster of Fig. \ref{GFRW counter-example}.  The application of Thm. \ref{sanchezthm} guarantees that this monster is geodesically complete for all geodesically complete spacelike foliations $\left( \Sigma , g_\Sigma \right)$.  Application of Eqs. \ref{eq:havg bgv}, \ref{eq:Havg asym past}, \ref{eq:Havg asym full} yield $H_{avg} \in \left( 0 , H \right]$ depending on choice of incomplete causal geodesic segment, $H^-_{avg} = 0$ for all past directed maximal geodesic rays, and $H^\pm_{avg} \in \left[ 0 , H \right]$ depending on parameterization and order the limits are taken over any maximal (both past and future) geodesic line chosen.  In particular, Thm. \ref{thm:bgv} may or may not apply, depending on the $H_{avg}$ variant chosen.  At this point we remind the reader that a strict reading of Thm. \ref{thm:bgv} cannot produce complete geodesic segments because all computations are performed over compact intervals.  However, with respect to $H^\pm_{avg}$ Thm. \ref{thm:bgv} both simultaneously does and does not apply depending on how the limit is taken (see Table I).

We now revisit Eq. \ref{eq:Havg sum}.  A natural decomposition of a maximal geodesic line into geodesic segments yields $\gamma = \gamma_{\left( -\infty , -1 \right]} \cup \gamma_{\left[ -1 , 0 \right]} \cup \gamma_{\left[ 0 , t_0 \right]} \cup \gamma_{\left[ t_0 , \infty \right)}$ where $\gamma_{\left( -\infty , -1 \right]}$ is the loitering asymptotic past, $\gamma_{\left[ -1 , 0 \right]}$ is the smooth transition region, $\gamma_{\left[ 0 , t_0 \right]}$ is the finite past de Sitter phase, and $\gamma_{\left[ t_0 , \infty \right)}$ is the de Sitter asymptotic future.  Ideally one would like the $H_{avg}$ sum of Eq. \ref{eq:Havg sum} to be equality.  However, from a mathematically pragmatic standpoint, what matters for the computation of $H_{avg}$ is the natural logarithm of the scale function at the boundary endpoints as is shown in \cite{Borde2003} Eq. 10, not what occurs in the interior.  This is one of the reasons Thm. \ref{thm:bgv} fails to predict geodesic completeness but Thm. \ref{sanchezthm} - which is derived by manually solving the geodesic equation - does.

The question now emerges, are such classical monsters physically reasonable?  In \cite{Mithani:2011en} the authors discuss the example of Fig. \ref{fig:eternally inflation}, namely $f\left( t \right) = a_0 \left( 1 + e^{H_0 t} \right)$.  Although not strictly an element of the above cohort of monsters due to the additive constant in the asymptotic future, the authors argue that such a scale function possesses non-zero quantum mechanical tunneling probability when treated with the Wheeler-DeWitt equation, thus eliminating the viability of an eternally loitering phase prior to inflation and expansion.  In particular, they argue that the class of aforementioned classical monsters (belonging to the ``emergent universe'' scenario \cite{Ellis:2002we,Ellis:2003qz,Mulryne:2005ef,delCampo:2011mq,Graham:2011nb}) should be excluded from the domain of discourse of physically reasonable scale factors (see also \cite{Aguirre:2013kea}).   In fact, the implications of cosmologies which infinitely loiter in a steady state generates the prime antinomy of \emph{temporal Copernicanism}: why do we exist NOW even though it is infinitely likely we should exist at some other time?  For the current discussion, however, we have little to contribute.

We do feel it is important to point out that calculational use of the Wheeler-DeWitt equation, predicated on the Arnowitt Deser and Misner (ADM) formalism --  see \cite{DeWitt:1967yk} and \cite{Arnowitt:1959ah} --  has somewhat of a contentious past, see \emph{e.\!\!~g.}~ \cite{Kiriushcheva2011,Landsman:1995gh}.  In particular, resolving the factor ordering problem eliminates the consistent canonical quantization of most general Hamiltonians in curved spacetimes~\cite{Dirac1964}, a detail which has simply been assumed to have been resolved in \cite{DeWitt:1967yk}.  With respect to the current problem at hand, one criticism of the calculation of \cite{Mithani:2011en} is the semi-classical treatment and subsequent elimination of time-ordering terms from the potential; a well known point of controversy apropos the Wheeler-DeWitt equation.  For future work, it would be interesting to repeat the calculation including these terms, possibly utilizing methods contained in \cite{Steigl2006}.  Many semi-classical physical quandaries are resolved when treated in a complete quantum formalism.  

Besides the above concern, it is not possible to have such a loitering universe exit the loitering phase or a bounce into an expanding phase in GR while obeying the weak energy condition (WEC): $T_{\mu\nu} t^\mu t^\nu \geq 0$, for timelike vectors $t^\sigma$.  As shown in Fig.~\ref{GFRW counter-example} the WEC is violated very briefly in the transition between the two phases during $t \in \left[ -1 , 0 \right]$.  Indeed, even the null energy condition (NEC) is violated in this region. However, this transition zone can be made arbitrarily small, perhaps even smaller than a Planck time, so the need for any such exotic matter such as galileons \cite{Nicolis:2008in} which can stably support such a NEC violation \cite{Creminelli:2006xe}, or any other new physics such as strings to accommodate a loitering phase \cite{Alexander:2000xv,Brandenberger:2001kj}, is fleeting. Of course, even the standard inflationary scenario necessarily violates the strong energy condition (SEC); indeed, our current period of accelerated expansion indicates the SEC is being violated even today, and it is well known that all of the energy conditions can be violated at the quantum level.

Because of their ability to violate the NEC, galileon fields have been used to construct monsters of the type discussed above, including loitering, inflationary and bouncing cosmologies \cite{Creminelli:2006xe,Creminelli:2010ba,Kobayashi:2010cm,Qiu:2011cy,Easson:2011zy,Cai:2012va}  and other exotic cosmological solutions \cite{Bains:2015gpv,Easson:2016klq,Easson:2018qgr}.
Stability arguments in such theories are complicated and depend on external background matter \cite{Easson:2013bda}. Even properly defining energy conditions in non-canonical scalar-tensor theories is non-trivial~\cite{Chatterjee:2012zh}. The devil is in the details, and it is unclear how similar quantum stability arguments of \cite{Mithani:2011en} apply to these models.

\section{Conclusion}
\label{sec:conclusion}
For a mathematical model to aspire to be a plausible description of the actual universe (or multiverse), additional criteria are often brought to bear. Such criteria may depend on the theoretical framework adopted in the treatment. Within classical theories of gravity, a frequently invoked question is whether a proposed solution is ‘generic’. More precisely, if the solution involves either a parameter value or initial conditions that must be chosen from a set of measure zero, the model is often dismissed as unphysical. A simple example is in Newtonian cosmology, where an initial density singularity will exist if the distribution of matter is exactly spherically symmetric. However, if the coefficients of the higher multipoles are not all strictly 0, infinite densities may not occur. In general relativity, however, departures from sphericity may not serve to remove the singularity if certain energy conditions apply, but it may be that most matter will ‘miss’ the singularity and the corresponding world lines will be geodesically complete in the past. In this paper, we have presented a set of geodesically complete solutions of inflationary metric theories of gravity that belong to an uncountable continuum, which is therefore a set of non-zero measure in the space of (continuous) initial conditions. As such, these ‘monsters’ are generic and thus are stable to perturbations in initial conditions, which are impossible to know precisely.  We note that the measure of the set of initial conditions being described here is predicated on the assumption of the continuity of spacetime, which may be transcended in the context of quantum cosmology.

Another criterion that might be used as a filter for physically plausible theories is the second law of thermodynamics, generalized to include cosmological event horizons. Such horizons are a feature of inflationary cosmological models. In the case of de Sitter space, the event horizon area is constant, but in FRW models where $f$ is concave upward as $t$ tends to infinity, the event horizon area will shrink and will generally imply a violation of the generalized second law of thermodynamics. Such models might then be regarded as unphysical. There may, of course be additional criteria beyond a solution being generic and not violating the generalized second law. The point we wish to make is that the plausibility or otherwise of a mathematical model to describe the real universe goes beyond it merely being a correct solution to a set of accepted dynamical equations.

In this paper, we offered a critical discussion of the $H_{avg} > 0$ geodesic completeness criterion for inflationary spacetimes with respect to the arguments presented in \cite{Borde2003}. Our first area of discussion involved the definition of inflationary spacetimes. In the voluminous body of literature on the subject, what is rigorously meant by an \emph{inflationary spacetime} varies between authors.  By introducing Defn.~\ref{my inflation defn}, we have suggested a standardized definition for inflation which underpins the geometrical requirements while broadly encompassing physically reasonable phenomenology.  We have further introduced $H^\gamma$ and $H^\gamma_{avg}$ and discussed both advantages and disadvantages to this definition, culminating with the generalizations of Eqs. \ref{eq:Havg asym past}, \ref{eq:Havg asym full} removing the underlying topological ambiguities of Thm. \ref{thm:bgv} by including the asymptotic past and future.  Drawing on this equation, we discussed an uncountably infinite cohort of classical monsters which
are geodesically complete despite having $H_{avg} > 0$ in some computations, in contradiction to the arguments found in~\cite{Borde2003} being supplanted by the calculations of \cite{Sanchez1994}.  These results are purely geometric and apply for any metrical theory of gravity. 

The solutions presented necessarily violated traditional energy conditions within the context of pure GR; although it is possible, such solutions may exist within non-canonical scalar field theories in a stable way as discussed above. Again within the context of GR the solutions may suffer from quantum considerations, as discussed in \cite{Mithani:2011en}, leading to the possible exclusion of this class of classical monsters from the domain of discourse for physically reasonable scale factors; however, a deeper understanding of quantum gravity is needed to definitively make such a statement.
For the time being, the question of the viability of eternal inflation, and the controversy of $H_{avg}$ is far from settled, leaving many possibilities for future work.

\section*{Acknowledgements} 
We are delighted to thank George Ellis,  Brett Kotschwar, Roger Penrose, and Tuna Yildirim for helpful discussions.  
DE is supported in part by the U.S. Department of Energy, Office of High Energy Physics, under Award No. DE-SC0019470, and the Foundational Questions Institute under Grant number FQXi-MGB-1927.  



\begin{thebibliography}{99}

\bibitem{HoyleNarlikar1964:2}
Hoyle F., Narlikar J. V. (1964) ``The C-Field as a Direct Particle Field''
Proceedings of the Royal Society of London. Series A, Mathematical and physical sciences, 1964-11-03, Vol.282 (1389), p.178-183

\bibitem{HoyleNarlikar1964:1}
Hoyle F., Narlikar J. V. (1964) ``On the Gravitational Influence of Direct Particle Fields''
Proceedings of the Royal Society of London. Series A, Mathematical and
Physical Sciences , Nov. 3, 1964, Vol. 282, No. 1389 (Nov. 3, 1964), pp. 184-190

\bibitem{HoyleNarlikar1966}
Hoyle F., Narlikar J. V. (1966) ``A Radical Departure from the `Steady State' Concept in Cosmology''
Proceedings of the Royal Society of London. Series A, Mathematical and physical sciences, 1966-02-22, Vol.290 (1421), p.162-176

\bibitem{Guth:1980zm}
  A.~H.~Guth,
  ``The Inflationary Universe: A Possible Solution To The Horizon And Flatness
  Problems,''
  Phys.\ Rev.\  D {\bf 23}, 347 (1981);
  
  A.~D.~Linde,
  ``A New Inflationary Universe Scenario: A Possible Solution Of The Horizon,
  Flatness, Homogeneity, Isotropy And Primordial Monopole Problems,''
  Phys.\ Lett.\  B {\bf 108}, 389 (1982);
  
  A.~Albrecht and P.~J.~Steinhardt,
  ``Cosmology For Grand Unified Theories With Radiatively Induced Symmetry
  Breaking,''
  Phys.\ Rev.\ Lett.\  {\bf 48}, 1220 (1982).
\bibitem{Linde:1986fc}
A.~D.~Linde,
``Eternal Chaotic Inflation,''
Mod. Phys. Lett. A \textbf{1}, 81 (1986)
doi:10.1142/S0217732386000129
\bibitem{Linde:1986fd}
A.~D.~Linde,
``Eternally Existing Selfreproducing Chaotic Inflationary Universe,''
Phys. Lett. B \textbf{175}, 395-400 (1986)
doi:10.1016/0370-2693(86)90611-8
\bibitem{Goncharov:1987ir}
A.~S.~Goncharov, A.~D.~Linde and V.~F.~Mukhanov,
``The Global Structure of the Inflationary Universe,''
Int. J. Mod. Phys. A \textbf{2}, 561-591 (1987)
doi:10.1142/S0217751X87000211

\bibitem{Borde:1993xh}
A.~Borde and A.~Vilenkin,
``Eternal inflation and the initial singularity,''
Phys. Rev. Lett. \textbf{72}, 3305-3309 (1994)
doi:10.1103/PhysRevLett.72.3305
[arXiv:gr-qc/9312022 [gr-qc]].

\bibitem{Borde2003}
A.~Borde, A.~H.~Guth and A.~Vilenkin,
``Inflationary space-times are incomplete in past directions,''
Phys. Rev. Lett. \textbf{90}, 151301 (2003)
doi:10.1103/PhysRevLett.90.151301
[arXiv:gr-qc/0110012 [gr-qc]].

\bibitem{Mithani:2012ii}
A.~Mithani and A.~Vilenkin,
``Did the universe have a beginning?,''
[arXiv:1204.4658 [hep-th]].

\bibitem{Hawking1975}
S.~W.~Hawking and G.~F.~R.~Ellis,
``The Large Scale Structure of Space-Time,''
doi:10.1017/CBO9780511524646

\bibitem{ONeill1983}
Barrett~O'Neill, 
``Semi-Riemannian Geometry with Applications to Relativity,"
 Academic Press, New York, NY, 1983.
 
\bibitem{Beem1996}
J.~K.~Beem, P.~Ehrlich and K.~Easley,
``Global Lorentzian Geometry,''
CRC Press, 1996.

\bibitem{Lakatos1976}
I.~Lakatos,
``Proofs and Refutations,''
Cambridge University Press, Port Melbourne, 1976.

\bibitem{Lee1997}
John M. Lee, 
``Riemannian Manifolds,"
 Graduate Texts in Mathematics, vol. 176, Springer New York, New York, NY, 1997.


\bibitem{Peterson2006}
Peter Peterson,, 
``Riemannian Geometry,"
 Graduate Texts in Mathematics, vol. 171, Springer New York, 2006.
 
\bibitem{Guth:2007ng}
A.~H.~Guth,
``Eternal inflation and its implications,''
J. Phys. A \textbf{40}, 6811-6826 (2007)
doi:10.1088/1751-8113/40/25/S25
[arXiv:hep-th/0702178 [hep-th]].
 
\bibitem{Liddle:2000cg}
A.~R.~Liddle and D.~H.~Lyth,
``Cosmological inflation and large scale structure'',
Cambridge University Press, 2000.

\bibitem{Bishop1969}
Bishop, B. L.,  O’Neill, B. (1969).
``Manifolds of Negative Curvature''
Trans. Amer. Math. Soc., 145, 1–49.
https://doi.org/10.1090/S0002-9947-1969-0251664-4
 
\bibitem{Sanchez1994}
Romero, A., Sanchez, M. (1994). On completeness of certain families of semi-Riemannian manifolds. 
Geometriae Dedicata, 53(1), 103–117. 
https://doi.org/10.1007/BF01264047

\bibitem{Sanchez1998}
Sanchez, M. (1998). On the Geometry of Generalized Robertson-Walker Spacetimes: Geodesics.
General Relativity and Gravitation, Vol. 30, No. 6, 915-932

\bibitem{Mithani:2011en}
A.~T.~Mithani and A.~Vilenkin,
``Collapse of simple harmonic universe,''
JCAP \textbf{01}, 028 (2012)
doi:10.1088/1475-7516/2012/01/028
[arXiv:1110.4096 [hep-th]].

\bibitem{Landsman:1995gh}
N.~P.~Landsman,
``Against the Wheeler-DeWitt equation,''
Class. Quant. Grav. \textbf{12}, L119-L124 (1995)
doi:10.1088/0264-9381/12/12/003
[arXiv:gr-qc/9510033 [gr-qc]].

\bibitem{Kiriushcheva2011}
Kiriushcheva, Natalia and Kuzmin, Sergei.
"The Hamiltonian formulation of general relativity: myths and reality" 
Open Physics, vol. 9, no. 3, 2011, pp. 576-615.
https://doi.org/10.2478/s11534-010-0072-2

\bibitem{DeWitt:1967yk}
B.~S.~DeWitt,
``Quantum Theory of Gravity. 1. The Canonical Theory,''
Phys. Rev. \textbf{160}, 1113-1148 (1967)
doi:10.1103/PhysRev.160.1113

\bibitem{Arnowitt:1959ah}
R.~L.~Arnowitt, S.~Deser and C.~W.~Misner,
``Dynamical Structure and Definition of Energy in General Relativity,''
Phys. Rev. \textbf{116}, 1322-1330 (1959)
doi:10.1103/PhysRev.116.1322

\bibitem{Steigl2006}
Šteigl, R., Hinterleitner F. 
``Factor ordering in standard quantum cosmology''
2006 Class. Quantum Grav. 23 3879

\bibitem{Dirac1964}
Dirac, P. A. M.,
``Lectures on Quantum Mechanics''
New York: Belfer Graduate School of Science, Yeshiva University (1964)

\bibitem{hetLam:2016llv}
H.~het Lam and T.~Prokopec,
``Singularities in FLRW Spacetimes,''
Phys. Lett. B \textbf{775}, 311-314 (2017)
doi:10.1016/j.physletb.2017.10.070
[arXiv:1606.01147 [gr-qc]].

\bibitem{Ellis:2002we}
G.~F.~R.~Ellis and R.~Maartens,
``The emergent universe: Inflationary cosmology with no singularity,''
Class. Quant. Grav. \textbf{21}, 223-232 (2004)
doi:10.1088/0264-9381/21/1/015
[arXiv:gr-qc/0211082 [gr-qc]].
\bibitem{Ellis:2003qz}
G.~F.~R.~Ellis, J.~Murugan and C.~G.~Tsagas,
``The Emergent universe: An Explicit construction,''
Class. Quant. Grav. \textbf{21}, no.1, 233-250 (2004)
doi:10.1088/0264-9381/21/1/016
[arXiv:gr-qc/0307112 [gr-qc]].
\bibitem{Mulryne:2005ef}
D.~J.~Mulryne, R.~Tavakol, J.~E.~Lidsey and G.~F.~R.~Ellis,
``An Emergent Universe from a loop,''
Phys. Rev. D \textbf{71}, 123512 (2005)
doi:10.1103/PhysRevD.71.123512
[arXiv:astro-ph/0502589 [astro-ph]].
\bibitem{delCampo:2011mq}
S.~del Campo, E.~I.~Guendelman, A.~B.~Kaganovich, R.~Herrera and P.~Labrana,
``Emergent Universe from Scale Invariant Two Measures Theory,''
Phys. Lett. B \textbf{699}, 211-216 (2011)
doi:10.1016/j.physletb.2011.03.061
[arXiv:1105.0651 [astro-ph.CO]].
\bibitem{Graham:2011nb}
P.~W.~Graham, B.~Horn, S.~Kachru, S.~Rajendran and G.~Torroba,
``A Simple Harmonic Universe,''
JHEP \textbf{02}, 029 (2014)
doi:10.1007/JHEP02(2014)029
[arXiv:1109.0282 [hep-th]].

\bibitem{Aguirre:2013kea}
A.~Aguirre and J.~Kehayias,
``Quantum Instability of the Emergent Universe,''
Phys. Rev. D \textbf{88}, 103504 (2013)
doi:10.1103/PhysRevD.88.103504
[arXiv:1306.3232 [hep-th]].

\bibitem{Nicolis:2008in}
A.~Nicolis, R.~Rattazzi and E.~Trincherini,
``The Galileon as a local modification of gravity,''
Phys. Rev. D \textbf{79}, 064036 (2009)
doi:10.1103/PhysRevD.79.064036
[arXiv:0811.2197 [hep-th]].

\bibitem{Creminelli:2006xe}
P.~Creminelli, M.~A.~Luty, A.~Nicolis and L.~Senatore,
``Starting the Universe: Stable Violation of the Null Energy Condition and Non-standard Cosmologies,''
JHEP \textbf{12}, 080 (2006)
doi:10.1088/1126-6708/2006/12/080
[arXiv:hep-th/0606090 [hep-th]].

\bibitem{Alexander:2000xv}
S.~Alexander, R.~H.~Brandenberger and D.~A.~Easson,
``Brane gases in the early universe,''
Phys. Rev. D \textbf{62}, 103509 (2000)
doi:10.1103/PhysRevD.62.103509
[arXiv:hep-th/0005212 [hep-th]].
\bibitem{Brandenberger:2001kj}
R.~Brandenberger, D.~A.~Easson and D.~Kimberly,
``Loitering phase in brane gas cosmology,''
Nucl. Phys. B \textbf{623}, 421-436 (2002)
doi:10.1016/S0550-3213(01)00636-8
[arXiv:hep-th/0109165 [hep-th]].

\bibitem{Creminelli:2010ba}
P.~Creminelli, A.~Nicolis and E.~Trincherini,
``Galilean Genesis: An Alternative to inflation,''
JCAP \textbf{11}, 021 (2010)
doi:10.1088/1475-7516/2010/11/021
[arXiv:1007.0027 [hep-th]].
\bibitem{Kobayashi:2010cm}
T.~Kobayashi, M.~Yamaguchi and J.~Yokoyama,
``G-inflation: Inflation driven by the Galileon field,''
Phys. Rev. Lett. \textbf{105}, 231302 (2010)
doi:10.1103/PhysRevLett.105.231302
[arXiv:1008.0603 [hep-th]].
\bibitem{Qiu:2011cy}
T.~Qiu, J.~Evslin, Y.~F.~Cai, M.~Li and X.~Zhang,
``Bouncing Galileon Cosmologies,''
JCAP \textbf{10}, 036 (2011)
doi:10.1088/1475-7516/2011/10/036
[arXiv:1108.0593 [hep-th]].
\bibitem{Easson:2011zy}
D.~A.~Easson, I.~Sawicki and A.~Vikman,
``G-Bounce,''
JCAP \textbf{11}, 021 (2011)
doi:10.1088/1475-7516/2011/11/021
[arXiv:1109.1047 [hep-th]].
\bibitem{Cai:2012va}
Y.~F.~Cai, D.~A.~Easson and R.~Brandenberger,
``Towards a Nonsingular Bouncing Cosmology,''
JCAP \textbf{08}, 020 (2012)
doi:10.1088/1475-7516/2012/08/020
[arXiv:1206.2382 [hep-th]].

\bibitem{Bains:2015gpv}
J.~S.~Bains, M.~P.~Hertzberg and F.~Wilczek,
``Oscillatory Attractors: A New Cosmological Phase,''
JCAP \textbf{05}, 011 (2017)
doi:10.1088/1475-7516/2017/05/011
[arXiv:1512.02304 [hep-th]].
\bibitem{Easson:2016klq}
D.~A.~Easson and A.~Vikman,
``The Phantom of the New Oscillatory Cosmological Phase,''
[arXiv:1607.00996 [gr-qc]].
\bibitem{Easson:2018qgr}
D.~A.~Easson and T.~Manton,
``Stable Cosmic Time Crystals,''
Phys. Rev. D \textbf{99}, no.4, 043507 (2019)
doi:10.1103/PhysRevD.99.043507
[arXiv:1802.03693 [hep-th]].

\bibitem{Easson:2013bda}
D.~A.~Easson, I.~Sawicki and A.~Vikman,
``When Matter Matters,''
JCAP \textbf{07}, 014 (2013)
doi:10.1088/1475-7516/2013/07/014
[arXiv:1304.3903 [hep-th]].

\bibitem{Chatterjee:2012zh}
S.~Chatterjee, D.~A.~Easson and M.~Parikh,
``Energy conditions in the Jordan frame,''
Class. Quant. Grav. \textbf{30}, 235031 (2013)
doi:10.1088/0264-9381/30/23/235031
[arXiv:1212.6430 [gr-qc]].

\bibitem{Nomizu1961}
Nomizu, K., Ozeki, H. (1961).  ``The existence of complete Riemannian metrics,''
Proc. Amer. Math. Soc. 12, 889

\bibitem{Minguzzi:2006sa}
E.~Minguzzi and M.~Sanchez,
``The Causal hierarchy of spacetimes,''
[arXiv:gr-qc/0609119 [gr-qc]].

\bibitem{Rado1962}
T. Rado, ``On non-computable functions," in The Bell System Technical Journal, vol. 41, no. 3, pp. 877-884, May 1962, doi: 10.1002/j.1538-7305.1962.tb00480.x.

\end{thebibliography}
\end{document}